\documentclass[fleqn,10pt]{wlscirep}

\usepackage[utf8]{inputenc}
\usepackage[T1]{fontenc}
\usepackage{mathtools}
\usepackage{bm}
\usepackage{graphicx}
\usepackage{hyperref}
\usepackage{color}
\definecolor{cy}{RGB}{0,205,205}   
\definecolor{teal}{HTML}{008080}

\newcommand{\sgr}{Sgr~$\rm{A}^*$\,}
\newcommand{\Msun}{M_{\odot}}

 \newcommand\aap{A\& A}%
 \newcommand\aj{AJ}%
 \newcommand\apj{ApJ}%
 \newcommand\mnras{MNRAS}%
 \newcommand\nat{Nature}%
 \newcommand\prd{Phys.Rev.D}%

\title{A Galactic centre gravitational-wave Messenger}

\author[1,2,3]{Marek Abramowicz}
\author[2,4]{Micha\l{} Bejger}
\author[5]{\'Eric Gourgoulhon}
\author[6,7*]{Odele Straub}
\affil[1]{Physics Department, University of Gothenburg, 412-96 G{\"o}teborg, Sweden}
\affil[2]{Nicolaus Copernicus Astronomical Center, Polish Academy of Sciences, 00-716, Warsaw, Poland}
\affil[3]{Physics Department, Silesian University of Opava, 74601 Opava, Czech Republic}
\affil[4]{APC, Universit\'e de Paris, CNRS/IN2P3, CEA/Irfu, Observatoire de Paris, 75013 Paris, France}
\affil[5]{LUTH, Observatoire de Paris, Universit\'e PSL, CNRS, Universit\'e de Paris, 5 place Jules Janssen, 92195 Meudon, France}
\affil[6]{Max Planck Institute for Extraterrestrial Physics, Giessenbachstr. 1, 85748, Garching, Germany}
\affil[7]{LESIA, Observatoire de Paris, Universit\'e PSL, CNRS, Universit\'e de Paris, 5 place Jules Janssen, 92195 Meudon, France}

\affil[*]{ostraub@mpe.mpg.de}

\begin{abstract}
Our existence in the Universe resulted from a rare combination of circumstances. The same must hold for any highly developed extraterrestrial civilisation, and if they have ever existed in the Milky Way, they would likely be scattered over large distances in space and time. However, all technologically advanced species must be aware of the unique property of the galactic centre: it hosts Sagittarius~$\rm{A}^*$ (\sgr), the closest supermassive black hole to anyone in the Galaxy. A civilisation with sufficient technical know-how may have placed material in orbit around \sgr for research, energy extraction, and communication purposes. In either case, its orbital motion will necessarily be a source of gravitational waves. We show that a Jupiter-mass probe on the retrograde innermost stable circular orbit around \sgr emits, depending on the black hole spin, at a frequency of $f_{GW} = 0.63 - 1.07 \,\mbox{mHz}$ and with a power of $P_{GW}=2.7\times\,10^{36} - 2.0\times\,10^{37}\,\mbox{erg/s}$. We discuss that the energy output of a single star is sufficient to stabilise the location of an orbiting probe for a billion years against gravitational wave induced orbital decay. Placing and sustaining a device near \sgr is therefore astrophysically possible. Such a probe will emit an unambiguously artificial continuous gravitational wave signal that is observable with LISA-type detectors.
\end{abstract}

\begin{document}

\flushbottom
\maketitle

\thispagestyle{empty}
%
\section*{Introduction} \label{sec-introduction}
%
It is conceivable that a civilisation advanced enough to see itself in a cosmological context would think of a way to make itself known to the Galaxy. We humans, for instance, sent in 1977 with Voyager 1 and 2 two postcards from Earth out into the vast depths of outer space. Destination: ``addressee unknown''.  The Voyagers carry not only instruments designed to study the planets in the outer Solar System, but also two golden records with messages intended for future humans or extraterrestrial lifeforms. Chances are that other civilisations would also try to announce their presence to the Galaxy. Using probes with instruments that emit radio waves aboard is one of the most obvious strategies to us. Indeed SETI, our most elaborated search programme for extraterrestrial intelligence, is based primarily on analysing radio waves \cite{2018AJ....156..260W}. The SETI Breakthrough Listen Initiative regularly releases data from surveys of the radio spectrum between 1 and 12 gigahertz (GHz) to the public. And we are not only searching -- on 17 November 1974 the largest radio antenna on Earth sent the famous ``Arecibo message'' to the globular cluster M13.

Of course, there is no particular reason to limit searching and announcing activities to radio waves. One may also think of signals encoded in the X-ray or $\gamma$-ray waves, or even in neutrinos. In addition, gravitational wave phenomena are omnipresent in the Universe. As opposed to electromagnetic waves, gravitational waves travel through space virtually undamped by matter along their way. In the case of the Advanced LIGO and Advanced Virgo detectors, waves emitted during the last stages of the inspiral and merger of binary systems of compact objects -- stellar mass black holes and neutron stars -- are firmly detectable from distances up to a few gigaparsec (Gpc, $1 pc = 3.086 \times 10^{18}$ cm) \cite{2018LRR....21....3A,2016PhRvL.116f1102A,2017PhRvL.118v1101A,2017PhRvL.119n1101A,2017PhRvL.119p1101A,ligovirgocat18}. 

A particular problem with searching for extraterrestrial civilisations is often summarised in Enrico Fermi's famous question {\it Where is everybody?}  \cite{fer+50}. Indeed, no signs of Aliens have ever been found. Why? Many physicists, cosmologists, and evolutionary biologists argue that a plausible answer to Fermi's question could be that our species {\it Homo Sapiens} resulted from an extremely rare combination of circumstances that started with the Big Bang, continued during the evolution of the Universe, the Galaxy and the Solar System, and proceeded during the Darwinian evolution on Earth. If the {\it rare Earth} principle \cite{war+00} is indeed the rule that limits the emergence of intelligent life, then advanced extraterrestrial civilisations should be genuinely rare -- scattered over vast distances in space and time. If the signal announcing somebody's existence originates always {\it in situ}, at  their respective planetary system, then all search strategies are condemned to be ``needle in the haystack'' searches. Any intrepid prospector must then cope with the omnipresent noise in which synthetic signals are buried. These signals have an {\it a priori unknown} physical nature, are characterised by {\it a priori unknown} frequencies and durations, and come from {\it a priori unknown} directions. Even if the rare Earth considerations will turn out to be irrelevant or incorrect, Fermi's wonderment -- Where is everybody? -- remains today a very relevant observational issue. No one had and has an obvious search strategy which would guarantee the success. Finding a signal requires not only skill but also luck.

We suggest a radically novel approach -- searching for a {\it unique, very particular signal} of an {\it a priori precisely and accurately known} nature, frequency and direction. Our suggestion follows from the hypothesis that once upon a time another civilisation existed in our Galaxy, a highly developed civilisation whose technolocial activities were only limited by the fundamental laws of Physics. They had at their disposal the expertise to operate on a grand interstellar scale, in particular the conversion of the power output of stars to run their instruments. Our suggestion may also apply to future humans with the necessary technical knowledge. If someone wanted to unambiguously announce their existence to the entire Galaxy, they would construct a ``Messenger'' with the following properties

\vskip0.05truecm
\vskip0.05truecm \noindent \hskip1.0truecm 1.~The Messenger {\bf location} is fundamentally unique and obvious to anyone in the Galaxy,
\vskip0.05truecm \noindent \hskip1.0truecm 2.~The {\bf physical nature} of the signal and the signal frequency are known a priori,
\vskip0.05truecm \noindent \hskip1.0truecm 3.~The {\bf emitted power} assures the signal's detectability in the whole Galaxy in terms of space,
\vskip0.05truecm \noindent \hskip1.0truecm 4.~Messenger's {\bf life time} assures the signal's detectability in the whole Galaxy in terms of time,
\vskip0.05truecm \noindent \hskip1.0truecm 5.~The {\bf energy supply} is provided by a natural astronomical phenomenon,
\vskip0.05truecm \noindent \hskip1.0truecm 6.~{\bf No maintenance} is needed; the Messenger is a fully autonomous device,
\vskip0.05truecm \noindent \hskip1.0truecm 7.~The {\bf artificial origin} of the signal unambiguously follows from its properties.
\vskip0.05truecm
\vskip0.05truecm 
\noindent We argue below that from fundamental laws of Physics and logical deduction it follows {\it necessarily} that such a Messenger should ideally be a Jupiter-mass black hole, orbiting \sgr for a few billion years at the retrograde innermost stable circular orbit (ISCO), and therefore naturally emitting gravitational waves with the frequency $f_{GW} = 0.63 - 1.07\,\mbox{mHz}$ and power $P_{GW}=2.7\times10^{36} - 2.0\times\,10^{37}\,$erg/s, depending on the black hole spin.

%
\section*{Location} \label{sec-Location}
%
%
Sagittarius~$\rm{A}^*$ (\sgr), the supermassive black hole in the centre of the Milky Way, is a unique object. Any advanced civilisation will, without any doubt, notice the existence of \sgr and recognise it as a unique location. We humans are aware of this for less than a century. Our recognition of \sgr started with the discovery of a radio hiss from the approximate direction of the Galactic centre \cite{jan33}, later X-ray emission from the same directions was detected during an early Aerobee survey \cite{bow+65}, but that it was a point source became clear only with radio interferometry \cite{bal+74}. It took a few decades more to identify this compact source as a black hole using infrared spectroscopy of the radial and proper motion of nearby stars \cite{gen+97,ghe+98}. Only recently \sgr has become our remote laboratory site where we can test gravity theories \cite{gravity18a,gravity18b}.

The mass of the Galactic centre black hole is very precisely known from direct measurements of stellar orbits around \sgr \cite{gravity18a}, $M = 4.14 \pm 0.03 \times 10^6 M_\odot$. The black hole spin $J$ is not known to date. Broadband spectral fitting indicates a broad range of the dimensionless Kerr spin parameter $a := J/M^2$, namely $a = 0.0 + 0.86$ (2$\sigma$ uncertainty) \cite{bro+11}. Once the black hole spin of \sgr can be determined, the gravitational wave frequency and power of any matter in orbit around it becomes very well constrained. 

The uniqueness of the location of the signal is guaranteed by placing the Messenger at a unequivocally distinguishable ``Keplerian'' orbit around \sgr. According to Einstein's general relativity, there is only one type of orbit to consider -- the innermost stable circular orbit (ISCO). All characteristics of the ISCO, in particular its radius, $r_0$, and the orbital frequency, $f_0$, depend only on the black hole mass, $M$, and its dimensionless spin, $a$. For non-rotating black holes, $a=0$, there is a single ISCO orbit, for $a \neq 0$ there is a pair of two well separated ISCO orbits in the equatorial plane: a closer ``prograde'' orbit (where the orbital momentum has the same direction like the angular momentum of the black hole) and a ``retrograde'' orbit (orbital momentum has the opposite direction to the angular momentum of the black hole). Retrograde orbits are indicated with a minus sign, e.g. $a=-0.9$. The best strategy would be to put the Messenger on the retrograde ISCO. Firstly, as we discuss later, the retrograde ISCO has a lower orbital energy than the prograde ISCO and secondly, natural astrophysical objects are less likely to settle on long lasting retrograde orbits than on prograde orbits. An object that stays for a long time at the retrograde ISCO must immediately be suspected to have an artificial origin.

\begin{figure}[h!]
\includegraphics[width=\textwidth]{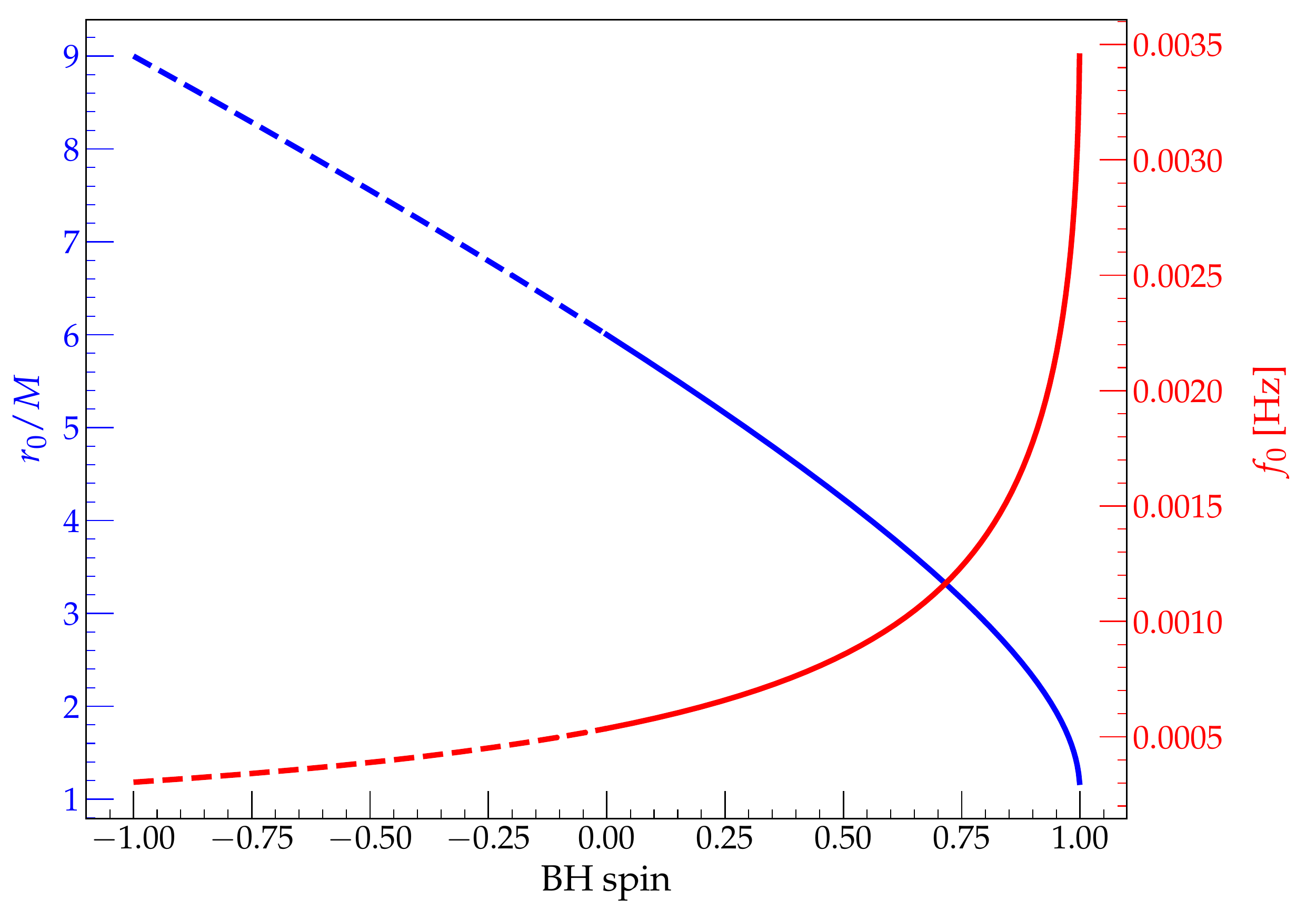}
\caption{
The ISCO radius $r_0$ (in blue) and ISCO orbital frequency $f_0$ (in red) as functions of the black hole spin $a$. The dashed lines mark the values for retrograde orbits. The ISCO radius is given in units of gravitational radius, $r_g = GM/c^2$ where $G=c=1$ and the frequency values are given for the mass of \sgr.}
\label{fig-isco-freq}
\end{figure}

Adopting for the mass the value $M = 4.1 \times 10^6\, M_\odot$ and for the spin $a=0$ and $\pm 0.9$, respectively, the radius $r_0$ (in Boyer-Lindquist coordinates) and the orbital frequency $f_0$ (as measured at infinity) at the prograde and retrograde ISCO radii are 

\begin{equation}
r_0  = \left\{\begin{array}{ll}
3.63 \times 10^{12} \,\mbox{cm} & \quad\mbox{for}\quad a={\color{white} +}\,0.0 \\
1.41 \times 10^{12} \,\mbox{cm} & \quad\mbox{for}\quad a=+\,0.9\\
5.27 \times 10^{12} \,\mbox{cm} & \quad\mbox{for}\quad a=-\,0.9
\end{array}\right.
~~~ \mbox{and} ~~~f_0  = \left\{\begin{array}{ll}
0.54 \,\mbox{mHz} & \quad\mbox{for}\quad a={\color{white} +}\,0.0 \\
1.78 \,\mbox{mHz} & \quad\mbox{for}\quad a=+\,0.9\\
0.32 \,\mbox{mHz} & \quad\mbox{for}\quad a=-\,0.9
\end{array}\right.
\label{eq-location-frequency-ISCO}  
\end{equation}
The functions $r_0 (a)$ and $f_0 (a)$ are shown in Figure \ref{fig-isco-freq}
%
%
Any circular orbital motion generates gravitational waves which are periodic with a dominant mode at twice the orbital frequency \cite{det+78,gou+18}. From equation (\ref{eq-location-frequency-ISCO}) we obtain the gravitational wave frequency of the dominant $m=2$ mode:
\begin{equation}
f_{\mbox{\scriptsize GW}} = 2 f_0 = \left\{\begin{array}{ll}
1.07 \,\mbox{mHz} & \quad\mbox{for}\quad a={\color{white} +}\,0.0 \\
3.55 \,\mbox{mHz} & \quad\mbox{for}\quad a=+\,0.9\\
0.63 \,\mbox{mHz} & \quad\mbox{for}\quad a=-\,0.9
\end{array}\right.
\label{eq-wave-frequency} 
\end{equation}
This frequency falls within the sensitivity range of the space borne gravitational-wave observatory LISA \cite{ama+17,rob+19}. The full waveform is described in \cite{gou+18}.

%
\section*{Physical nature. Emitted power} \label{sec-nature-power}
%
In order to generate a reasonably strong gravitational-wave signal, the body orbiting \sgr must have a mass in an astronomical range. It cannot be too small or it does not produce a gravitational-wave amplitude detectable over a substantial distance for \sgr, and it cannot be too big or it will cost too much energy to sustain its orbit over a sufficiently long time. Our calculations indicate that a detectable Messenger mass is in the ballpark of moons, planets, or stars.

The energy loss due to gravitational radiation by an object of mass $m$ in a \textit{circular orbit} around a rotating black hole of mass $M \gg m$ has been computed in \cite{gou+18} (The result can be found in the public \href{https://cocalc.com/share/11745bc1-4f1c-4eb8-ae69-3a4d90eeb6b6/Messenger_power.ipynb?viewer=share}{SageMath notebook} listed at the end of the article). For a Jupiter-mass Messenger, $m = 2 \times 10^{27}$ kg, orbiting at ISCO [Eq.~(\ref{eq-location-frequency-ISCO})], this yields the following numerical values of the total radiated power:
\begin{equation}
P_{\mbox{\scriptsize GW}} = \frac{\mathrm{d} E}{\mathrm{d} t} =
\left\{\begin{array}{ll}
2.0 \times 10^{37}  \; {\rm erg\; s}^{-1} & \quad\mbox{for}\quad   a={\color{white} +}\,0.0 \\
7.8 \times 10^{38}  \; {\rm erg\; s}^{-1} & \quad\mbox{for}\quad   a=+\,0.9 \\
2.7 \times 10^{36}  \; {\rm erg\; s}^{-1} & \quad\mbox{for}\quad  a=-\,0.9
\end{array}\right.
\label{eq-power-emitted} 
\end{equation}
Values of $P_{\mbox{\scriptsize GW}}$ for other values of $m$ can be found in Table~1. The above results relies on exact computations in the Kerr metric \cite{gou+18}. As a check, let us note that an approximate estimate of $P_{\mbox{\scriptsize GW}}$ can be obtained by a 1.5-order post-Newtonian formula based on \cite{shi94}:
\begin{equation}
 P_{\mbox{\scriptsize GW}} \simeq \frac{32}{5}\left(\frac{M}{r_0}\right)^5 \left(\frac{m}{M}\right)^2 \times \left( 1 + \frac{13}{168}\frac{M}{2 r_0} - \frac{73}{12} \frac{a M^3}{L^3}  \right) \times \frac{c^5}{G} ,
\label{eq-power-post-Newton} 
\end{equation}
where $L$ the specific angular momentum of a Keplerian orbit. Equation~(\ref{eq-power-post-Newton}) results in values in agreement with Eq.~(\ref{eq-power-emitted}) up to $\sim 10\%$.
%
%
\begin{table*}[h]
\centering
\renewcommand{\arraystretch}{1.5}
\begin{tabular}{| l | c | c | c |}
\hline
$a$                     & 0          & 0.9        &--\,0.9                        \\
\hline
$P\left(m_{\mbox{\scriptsize Sun}}\right)$     & $ 10^{43}$  & $ 10^{45}$       &  $  10^{42}$     \\
$P\left(m_{\mbox{\scriptsize Jupiter}}\right)$ & $ 10^{37}$ & $ 10^{39}$        &  $ 10^{36}$      \\
$P\left(m_{\mbox{\scriptsize Earth}}\right)$   & $ 10^{31}$ & $ 10^{33}$        &  $ 10^{30}$     \\
\hline
\end{tabular}
\caption{Power required to hold the Messenger on the innermost stable circular orbit,
in ${\rm erg\; s}^{-1}$. This corresponds to the gravitational-wave power, $P_{\mbox{\scriptsize GW}} = {\mathrm{d}E}/{\mathrm{d}t}$ , calculated in Eq.~(\ref{eq-power-emitted}), for a given point mass $m$ orbiting a massive black hole, of mass $M$ and spin $a$.}
\label{tab:dEdt}
\end{table*}

The galactic centre is a busy place with a nuclear star cluster, dense cloudy objects, and a putative population of stellar remnants. At the location of the ISCO, in close proximity to the black hole, the environment consists mostly of ionised gas. Jupiter as such would be tidally disrupted at the ISCO. The same is valid for stars. Only objects with densities $>10^6\; {\rm g\; cm}^{-3}$ (comparable to a white dwarf) are safe \cite{gou+18}. A conceivable strategy have a ``Jupiter'' at the ISCO would be to compress it, or collapse it into a black hole. A discussion of the engineering aspects would go beyond the scope of our work. Since only small object like brown dwarfs, planets, and asteroids survive the tidal force of \sgr we don't expect any major dynamical encounters with a satellite at the ISCO. The probability of a rock having a close encounter with a Jupiter mass black hole - which has a radius of less than 300cm - is negligibly small. The resident gas can exert a drag force on the Messenger. The power of the drag force is given by the density of the gas as well as the velocity and area of the orbiting body. In the specific case of a Jupiter-mass black hole in orbit around \sgr, which is in all probability surrounded by a very tenuous plasma, $P_{\mbox{\scriptsize drag}} \propto \rho v^3 A \approx 10^{18} \; {\rm erg\; s}^{-1}$, is negligible with respect to Eq.~(\ref{eq-power-emitted})

For a mass $m = 2 \times 10^{27}$~kg orbiting at the ISCO, it has been calculated in Ref.~\cite{gou+18} that within $T_{\rm obs} = 1$~yr of observational time, LISA can detect the gravitational waves with a signal-to-noise ratio as high as
\begin{equation}
\mbox{SNR}_{\mbox{\scriptsize GW}} = \left\{\begin{array}{ll}
2.8 \times 10^2 & \quad\mbox{for}\quad a={\color{white} +}\,0.0 \\
3.6 \times 10^3 & \quad\mbox{for}\quad a=+\,0.9\\
5.5 \times 10^1 & \quad\mbox{for}\quad a=-\,0.9
\label{eq-signal-to-noise} 
\end{array}\right.
\end{equation}
Such a signal will be easily recognised as a true gravitational-wave emission. Within an observational period of five years, LISA can detect orbiter masses as small as super-Earths.

%
\section*{Lifetime. Artificial origin} \label{sec-lifetime-origin}
%
Nobody knows how long a technically-advanced civilisation lives, but it is likely that their lifetime is significantly shorter than a few billion years. Therefore, the Messenger beacon should last for about this time
\begin{equation}
\mbox{The Messenger lifetime:}\,~ t_{\mbox{\scriptsize life}} \approx 10^9 \; {\rm yr} = 10^{17} \; {\rm s}
\label{eq-life-time} 
\end{equation}
otherwise its signal could be easily missed by other Galactic civilisations. Thus, the lifetime of the Messenger will be much longer than that of the civilisation that created it. If we ever discover the gravitational signal discussed here, it will possibly be a message from a dead civilisation. Loeb \cite{loe18}  has discussed this issue in a different context of searching for ``space junk'' -- relics of dead civilisations.

The power [Eq.~(\ref{eq-power-emitted})] must be continuously supplied to the Messenger in order to compensate the orbital energy losses due to gravitational radiation. Without this supply, the Messenger would drift out of the orbit and plunge into the black hole \sgr. The change of the orbital frequency in time, $\mathrm{d} f/\mathrm{d} t$ is diverging at the ISCO \cite{fin+00,gou+18}, as indicated in Fig.~2. For comparison, long-lasting observations of LISA provide the following frequency derivative resolution: if the observation time of LISA is $T_{\rm obs}$, the frequency resolution (width of the frequency bin) is $\delta f = 1/T_{\rm obs}$. Minimal frequency derivative $\dot{f} = \mathrm{d} f / \mathrm{d} t$ which can be detected (signals with smaller $\dot{f}$ will stay within the same frequency bin in time $T_{\rm obs}$) is then $\dot{f}_{\rm min} = \delta f / T_{\rm obs} = 1/T_{\rm obs}^2$. For $T_{\rm obs}$ = 1 year, $\dot{f}_{\rm min} \simeq 10^{-15}$ Hz/s, which, in view of Fig.~2, is more than enough to distinguish the signal from a naturally decaying orbit at ISCO from a strictly monochromatic synthetic signal.

\begin{figure}[h!]
\includegraphics[width=\textwidth]{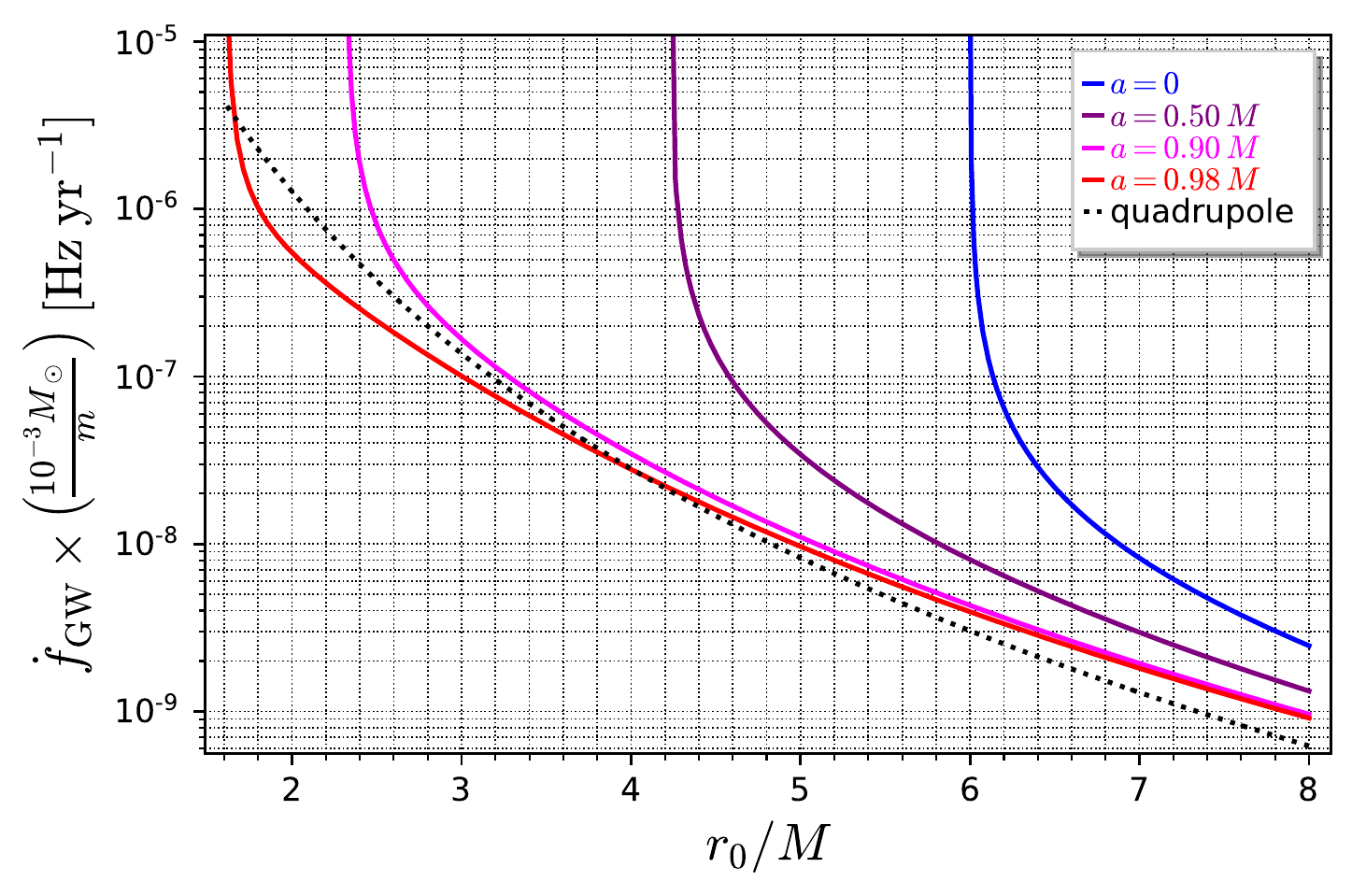}
\caption{
Secular frequency change due to gravitational radiation loss in terms of the orbital radius $r_0$, as computed in \cite{gou+18}. The corresponding \href{https://cocalc.com/share/11745bc1-4f1c-4eb8-ae69-3a4d90eeb6b6/frequency_change.ipynb?viewer=share}{SageMath notebook} is publicly available. At the ISCO the total energy of the orbiting body $E = E(r_0)$ is minimal so that $\mathrm{d} E/\mathrm{d} r_0 = 0$. This implies a steep increase of the frequency derivative and $\mathrm{d} f_0/\mathrm{d} t = +\infty$ at the ISCO.}
\label{fig:df}
\end{figure}
This is in our opinion the most important feature of the Messenger concept. If an observer detects a continuous, long-lasting (of the order of months or years), gravitational-wave signal with a stable frequency corresponding to the retrograde ISCO of \sgr, then it is obvious that this signal is artificial.

%
\section*{Energy supply. No maintenance } \label{sec-supply}
%
Given Eq.~(\ref{eq-power-emitted}), the total energy needed to supply a Jupiter-mass Messenger over its lifetime is
\begin{equation}
{\cal E} = t_{\mbox{\scriptsize life}} \, P_{\mbox{\scriptsize GW}} \approx
\left\{\begin{array}{ll}
2.0 \times 10^{54}  \; {\rm erg\;} & \quad\mbox{for}\quad   a={\color{white} +}\,0.0 \\
7.8 \times 10^{55}  \; {\rm erg\;} & \quad\mbox{for}\quad   a=+\,0.9 \\
2.7 \times 10^{53}  \; {\rm erg\;} & \quad\mbox{for}\quad   a=-\,0.9
\end{array}\right.
\label{eq-energy-neede} 
\end{equation}
Restricted to the retrograde ISCO, this corresponds to a fuel mass ${\cal M} = {\cal E}/c^2 \sim \; 3\times10^{32} \; \mbox{--} \;\; 2\times10^{33} \; {\rm g}$, depending on the black hole spin.

Therefore, the mass of the Messenger {\it cannot} itself  be a source for the energy supply needed even in the extreme case of the 100\% efficiency of the $E=mc^2$ conversion -- energy has to be supplied from outside. Fortunately, a natural process involving a single star of about $0.1 - 1 \Msun$, will suffice if the Messenger mass is of the order of a Jupiter mass. For a Messenger with a mass as much as  $1 \Msun$, however, the energy needed to sustain its orbit would be 6 orders of magnitude larger, meaning that a supply by one star would be not possible.

\section*{Discussion} \label{sec-Discussion}
%
The ``Galactic Centre Messenger'' is a thought experiment that originated from the question, what kind of a mass on what kind of an orbit around \sgr would produce a measurable gravitational wave signal and what kind of an energy would be required to make that signal continuous. We show by means of a few simple calculations that the energy supply of one solar mass can sustain one Jupiter mass in an orbit for one billion years. This means it is energetically feasible to stabilise a gravitational wave emitting mass for a few billion years in orbit close to the Galactic centre black hole. Given this finding, we want to propose gravitational waves as a new road to detect, or search for, intelligent life in the Galaxy. It is a promising and exciting way that could be used in conjunction with the classical SETI radio searches, much like we perform multi-messenger astronomy with LIGO/Virgo and electromagnetic observatories today.

Would future humans or highly developed extraterrestrial civilisations actually position a device with a Messenger function in orbit around \sgr?  Nobody knows their intentions or technological skills. They might construct a Jupiter-mass probe at the ISCO to explore the supermassive black hole in the Galactic centre, or to extract and harness its energy, or even for intentions unfathomable to the human mind. Such devices would also serve as Messengers and their gravitational wave signature may be picked up by LISA. Alternatively, they might choose to assemble a radio (or other electromagnetic wave) transmitter. Gravitational and electromagnetic wave beacons may a priori be located anywhere. The easiest and cheapest way would perhaps be to mount an antenna on a satellite around a stellar remnant in ones neighbourhood. But even the brightest electromagnetic signal is easily missed because its direction and wavelength are unknown to others. We argue that if someone wanted to communicate in the Galaxy, which is an intrinsically difficult task given the Galactic distances and morphology that have to be overcome, the Galactic centre black hole is a predictable focal point, i.e. the Schelling point of our Galaxy. Schelling asked \cite{sch60}: ``If you are to meet a stranger in New York City, but you cannot communicate with the person, then when and where will you choose to meet?'' Similarly, in a Galaxy where it is hard to communicate, we ask, where will you choose to look? The most natural answer is \sgr. 

We do not know ``where everybody is'' but we know one thing for sure: already in the first year of its operation LISA will be able to verify if a Jupiter-mass orbiter is present in the Galactic centre. Smaller masses will require longer monitoring to reach a conclusive signal-to-noise ratio. The absence of a continuous signal from the direction of \sgr does, of course, not imply that extraterrestrial life has not evolved elsewhere. A successful detection, however, will provide a definite and unambiguous \textit{proof} that an intelligent civilisation did exist in our Galaxy.

\section*{Data availability}
\textbf{Accession codes} SageMath notebooks: \\
{\footnotesize \url{https://cocalc.com/share/11745bc1-4f1c-4eb8-ae69-3a4d90eeb6b6/frequency_change.ipynb?viewer=share}}\\
{\footnotesize \url{https://cocalc.com/share/11745bc1-4f1c-4eb8-ae69-3a4d90eeb6b6/Messenger_power.ipynb?viewer=share}}


\begin{thebibliography}{10}
\urlstyle{rm}
\expandafter\ifx\csname url\endcsname\relax
  \def\url#1{\texttt{#1}}\fi
\expandafter\ifx\csname urlprefix\endcsname\relax\def\urlprefix{URL }\fi
\expandafter\ifx\csname doiprefix\endcsname\relax\def\doiprefix{DOI: }\fi
\providecommand{\bibinfo}[2]{#2}
\providecommand{\eprint}[2][]{\url{#2}}

\bibitem{2018AJ....156..260W}
\bibinfo{author}{{Wright}, J.~T.}, \bibinfo{author}{{Kanodia}, S.} \&
  \bibinfo{author}{{Lubar}, E.}
\newblock \bibinfo{journal}{\bibinfo{title}{{How Much SETI Has Been Done?
  Finding Needles in the n-dimensional Cosmic Haystack}}}.
\newblock {\emph{\JournalTitle{\aj}}} \textbf{\bibinfo{volume}{156}},
  \bibinfo{pages}{260}, \doiprefix\url{10.3847/1538-3881/aae099}
  (\bibinfo{year}{2018}).
\newblock \eprint{1809.07252}.

\bibitem{2018LRR....21....3A}
\bibinfo{author}{{Abbott}, B.~P.} \emph{et~al.}
\newblock \bibinfo{journal}{\bibinfo{title}{{Prospects for observing and
  localizing gravitational-wave transients with Advanced LIGO, Advanced Virgo
  and KAGRA}}}.
\newblock {\emph{\JournalTitle{Living Reviews in Relativity}}}
  \textbf{\bibinfo{volume}{21}}, \bibinfo{pages}{3},
  \doiprefix\url{10.1007/s41114-018-0012-9} (\bibinfo{year}{2018}).
\newblock \eprint{1304.0670}.

\bibitem{2016PhRvL.116f1102A}
\bibinfo{author}{{Abbott}, B.~P.} \emph{et~al.}
\newblock \bibinfo{journal}{\bibinfo{title}{{Observation of Gravitational Waves
  from a Binary Black Hole Merger}}}.
\newblock {\emph{\JournalTitle{Physical Review Letters}}}
  \textbf{\bibinfo{volume}{116}}, \bibinfo{pages}{061102},
  \doiprefix\url{10.1103/PhysRevLett.116.061102} (\bibinfo{year}{2016}).
\newblock \eprint{1602.03837}.

\bibitem{2017PhRvL.118v1101A}
\bibinfo{author}{{Abbott}, B.~P.} \emph{et~al.}
\newblock \bibinfo{journal}{\bibinfo{title}{{GW170104: Observation of a
  50-Solar-Mass Binary Black Hole Coalescence at Redshift 0.2}}}.
\newblock {\emph{\JournalTitle{Physical Review Letters}}}
  \textbf{\bibinfo{volume}{118}}, \bibinfo{pages}{221101},
  \doiprefix\url{10.1103/PhysRevLett.118.221101} (\bibinfo{year}{2017}).
\newblock \eprint{1706.01812}.

\bibitem{2017PhRvL.119n1101A}
\bibinfo{author}{{Abbott}, B.~P.} \emph{et~al.}
\newblock \bibinfo{journal}{\bibinfo{title}{{GW170814: A Three-Detector
  Observation of Gravitational Waves from a Binary Black Hole Coalescence}}}.
\newblock {\emph{\JournalTitle{Physical Review Letters}}}
  \textbf{\bibinfo{volume}{119}}, \bibinfo{pages}{141101},
  \doiprefix\url{10.1103/PhysRevLett.119.141101} (\bibinfo{year}{2017}).
\newblock \eprint{1709.09660}.

\bibitem{2017PhRvL.119p1101A}
\bibinfo{author}{{Abbott}, B.~P.} \emph{et~al.}
\newblock \bibinfo{journal}{\bibinfo{title}{{GW170817: Observation of
  Gravitational Waves from a Binary Neutron Star Inspiral}}}.
\newblock {\emph{\JournalTitle{Physical Review Letters}}}
  \textbf{\bibinfo{volume}{119}}, \bibinfo{pages}{161101},
  \doiprefix\url{10.1103/PhysRevLett.119.161101} (\bibinfo{year}{2017}).
\newblock \eprint{1710.05832}.

\bibitem{ligovirgocat18}
\bibinfo{author}{{The LIGO Scientific Collaboration}} \& \bibinfo{author}{{the
  Virgo Collaboration}}.
\newblock \bibinfo{journal}{\bibinfo{title}{{GWTC-1: A Gravitational-Wave
  Transient Catalog of Compact Binary Mergers Observed by LIGO and Virgo during
  the First and Second Observing Runs}}}.
\newblock {\emph{\JournalTitle{Physical Review X}}}
  \textbf{\bibinfo{volume}{9}}, \bibinfo{pages}{031040},
  \doiprefix\url{10.1103/PhysRevX.9.031040} (\bibinfo{year}{2019}).
\newblock \eprint{1811.12907}.

\bibitem{fer+50}
\bibinfo{author}{Fermi, E.}
\newblock \bibinfo{title}{A lunch conversation with edward teller, herbert york
  and emil konopinski.} (\bibinfo{year}{1950}).

\bibitem{war+00}
\bibinfo{author}{{Ward}, P.~D.}, \bibinfo{author}{{Brownlee}, D.} \&
  \bibinfo{author}{{Krauss}, L.}
\newblock \bibinfo{journal}{\bibinfo{title}{{Rare Earth: Why Complex Life Is
  Uncommon in the Universe}}}.
\newblock {\emph{\JournalTitle{Physics Today}}} \textbf{\bibinfo{volume}{53}},
  \bibinfo{pages}{62}, \doiprefix\url{10.1063/1.1325239}
  (\bibinfo{year}{2000}).

\bibitem{jan33}
\bibinfo{author}{{Jansky}, K.~G.}
\newblock \bibinfo{journal}{\bibinfo{title}{{Radio Waves from Outside the Solar
  System}}}.
\newblock {\emph{\JournalTitle{\nat}}} \textbf{\bibinfo{volume}{132}},
  \bibinfo{pages}{66}, \doiprefix\url{10.1038/132066a0} (\bibinfo{year}{1933}).

\bibitem{bow+65}
\bibinfo{author}{{Bowyer}, S.}, \bibinfo{author}{{Byram}, E.~T.},
  \bibinfo{author}{{Chubb}, T.~A.} \& \bibinfo{author}{{Friedman}, H.}
\newblock \bibinfo{journal}{\bibinfo{title}{{Cosmic X-ray Sources}}}.
\newblock {\emph{\JournalTitle{Science}}} \textbf{\bibinfo{volume}{147}},
  \bibinfo{pages}{394--398}, \doiprefix\url{10.1126/science.147.3656.394}
  (\bibinfo{year}{1965}).

\bibitem{bal+74}
\bibinfo{author}{{Balick}, B.} \& \bibinfo{author}{{Brown}, R.~L.}
\newblock \bibinfo{journal}{\bibinfo{title}{{Intense sub-arcsecond structure in
  the galactic center}}}.
\newblock {\emph{\JournalTitle{\apj}}} \textbf{\bibinfo{volume}{194}},
  \bibinfo{pages}{265--270}, \doiprefix\url{10.1086/153242}
  (\bibinfo{year}{1974}).

\bibitem{gen+97}
\bibinfo{author}{{Genzel}, R.}, \bibinfo{author}{{Eckart}, A.},
  \bibinfo{author}{{Ott}, T.} \& \bibinfo{author}{{Eisenhauer}, F.}
\newblock \bibinfo{journal}{\bibinfo{title}{{On the nature of the dark mass in
  the centre of the Milky Way}}}.
\newblock {\emph{\JournalTitle{\mnras}}} \textbf{\bibinfo{volume}{291}},
  \bibinfo{pages}{219--234}, \doiprefix\url{10.1093/mnras/291.1.219}
  (\bibinfo{year}{1997}).

\bibitem{ghe+98}
\bibinfo{author}{{Ghez}, A.~M.}, \bibinfo{author}{{Klein}, B.~L.},
  \bibinfo{author}{{Morris}, M.} \& \bibinfo{author}{{Becklin}, E.~E.}
\newblock \bibinfo{journal}{\bibinfo{title}{{High Proper-Motion Stars in the
  Vicinity of Sagittarius A*: Evidence for a Supermassive Black Hole at the
  Center of Our Galaxy}}}.
\newblock {\emph{\JournalTitle{\apj}}} \textbf{\bibinfo{volume}{509}},
  \bibinfo{pages}{678--686} (\bibinfo{year}{1998}).
\newblock \eprint{astro-ph/9807210}.

\bibitem{gravity18a}
\bibinfo{author}{{GRAVITY Collaboration}} \emph{et~al.}
\newblock \bibinfo{journal}{\bibinfo{title}{{Detection of the gravitational
  redshift in the orbit of the star S2 near the Galactic centre massive black
  hole}}}.
\newblock {\emph{\JournalTitle{\aap}}} \textbf{\bibinfo{volume}{615}},
  \bibinfo{pages}{L15}, \doiprefix\url{10.1051/0004-6361/201833718}
  (\bibinfo{year}{2018}).
\newblock \eprint{1807.09409}.

\bibitem{gravity18b}
\bibinfo{author}{{GRAVITY Collaboration}} \emph{et~al.}
\newblock \bibinfo{journal}{\bibinfo{title}{{Detection of orbital motions near
  the last stable circular orbit of the massive black hole SgrA*}}}.
\newblock {\emph{\JournalTitle{\aap}}} \textbf{\bibinfo{volume}{618}},
  \bibinfo{pages}{L10}, \doiprefix\url{10.1051/0004-6361/201834294}
  (\bibinfo{year}{2018}).
\newblock \eprint{1810.12641}.

\bibitem{bro+11}
\bibinfo{author}{{Broderick}, A.~E.}, \bibinfo{author}{{Fish}, V.~L.},
  \bibinfo{author}{{Doeleman}, S.~S.} \& \bibinfo{author}{{Loeb}, A.}
\newblock \bibinfo{journal}{\bibinfo{title}{{Evidence for Low Black Hole Spin
  and Physically Motivated Accretion Models from Millimeter-VLBI Observations
  of Sagittarius A*}}}.
\newblock {\emph{\JournalTitle{\apj}}} \textbf{\bibinfo{volume}{735}},
  \bibinfo{pages}{110}, \doiprefix\url{10.1088/0004-637X/735/2/110}
  (\bibinfo{year}{2011}).
\newblock \eprint{1011.2770}.

\bibitem{det+78}
\bibinfo{author}{{Detweiler}, S.~L.}
\newblock \bibinfo{journal}{\bibinfo{title}{{Black holes and gravitational
  waves. I. Circular orbits about a rotating hole.}}}
\newblock {\emph{\JournalTitle{\apj}}} \textbf{\bibinfo{volume}{225}},
  \bibinfo{pages}{687--693}, \doiprefix\url{10.1086/156529}
  (\bibinfo{year}{1978}).

\bibitem{gou+18}
\bibinfo{author}{{Gourgoulhon}, E.}, \bibinfo{author}{{Le Tiec}, A.},
  \bibinfo{author}{{Vincent}, F.~H.} \& \bibinfo{author}{{Warburton}, N.}
\newblock \bibinfo{journal}{\bibinfo{title}{{Gravitational waves from bodies
  orbiting the Galactic center black hole and their detectability by LISA}}}.
\newblock {\emph{\JournalTitle{\aap}}} \textbf{\bibinfo{volume}{627}},
  \bibinfo{pages}{A92}, \doiprefix\url{10.1051/0004-6361/201935406}
  (\bibinfo{year}{2019}).
\newblock \eprint{1903.02049}.

\bibitem{ama+17}
\bibinfo{author}{{Amaro-Seoane}, P.} \emph{et~al.}
\newblock \bibinfo{journal}{\bibinfo{title}{{Laser Interferometer Space
  Antenna}}}.
\newblock {\emph{\JournalTitle{arXiv e-prints}}}  (\bibinfo{year}{2017}).
\newblock \eprint{1702.00786}.

\bibitem{rob+19}
\bibinfo{author}{{Robson}, T.}, \bibinfo{author}{{Cornish}, N.~J.} \&
  \bibinfo{author}{{Liu}, C.}
\newblock \bibinfo{journal}{\bibinfo{title}{{The construction and use of LISA
  sensitivity curves}}}.
\newblock {\emph{\JournalTitle{Classical and Quantum Gravity}}}
  \textbf{\bibinfo{volume}{36}}, \bibinfo{pages}{105011},
  \doiprefix\url{10.1088/1361-6382/ab1101} (\bibinfo{year}{2019}).
\newblock \eprint{1803.01944}.

\bibitem{shi94}
\bibinfo{author}{{Shibata}, M.}
\newblock \bibinfo{journal}{\bibinfo{title}{{Gravitational waves by compact
  star orbiting around rotating supermassive black holes}}}.
\newblock {\emph{\JournalTitle{\prd}}} \textbf{\bibinfo{volume}{50}},
  \bibinfo{pages}{6297--6311}, \doiprefix\url{10.1103/PhysRevD.50.6297}
  (\bibinfo{year}{1994}).

\bibitem{loe18}
\bibinfo{author}{{Loeb}, A.}
\newblock \bibinfo{journal}{\bibinfo{title}{{How to Search for Dead Cosmic
  Civilizations}}}.
\newblock {\emph{\JournalTitle{Scientific American}}}  (\bibinfo{year}{2018}).

\bibitem{fin+00}
\bibinfo{author}{{Finn}, L.~S.} \& \bibinfo{author}{{Thorne}, K.~S.}
\newblock \bibinfo{journal}{\bibinfo{title}{{Gravitational waves from a compact
  star in a circular, inspiral orbit, in the equatorial plane of a massive,
  spinning black hole, as observed by LISA}}}.
\newblock {\emph{\JournalTitle{\prd}}} \textbf{\bibinfo{volume}{62}},
  \bibinfo{pages}{124021}, \doiprefix\url{10.1103/PhysRevD.62.124021}
  (\bibinfo{year}{2000}).
\newblock \eprint{gr-qc/0007074}.

\bibitem{sch60}
\bibinfo{author}{{Schelling}, T.~C.}
\newblock \emph{\bibinfo{title}{The strategy of conflict}}
  (\bibinfo{publisher}{Cambridge University Press}, \bibinfo{year}{1960}).

\end{thebibliography}

\section*{Acknowledgements}

M.A. acknowledges the Polish NCN grant 2015/19/B/ST9/01099 and the Czech Science Foundation grant No.~17-16287S which supported his visits to Paris Observatory. M.B. was partially supported by the Polish NCN grant 2016/22/E/ST9/00037.

\section*{Author contributions statement}
Idea: M.A. and E.G.; Text: O.S. and M.A.; Calculations: E.G., O.S., M.A., and M.B.; Figures: O.S. and E.G. All authors contributed equally to the discussion and review of the manuscript.

\section*{Additional information}
\textbf{Competing interests} The authors declare no competing interests.

\end{document}